\documentstyle[epsfig]{aipproc}

\begin{document}
\title{A Binary Neutron Star GRB Model}

\author{J.\ R.\ Wilson$^{*\dagger}$, J.\ D.\ Salmonson$^*$, \& G.\ J.\
Mathews$^{\dagger}$} 

\address{$^*$University of California, Lawrence Livermore National
Laboratory,\\ Livermore, California 94550\\ $^{\dagger}$University of
Notre Dame, Department of Physics, Notre Dame, Indiana 46556}

\maketitle

\begin{abstract}
In this paper we present the preliminary results of a model for the
production of gamma-ray bursts (GRBs) through the compressional
heating of binary neutron stars near their last stable orbit prior to
merger.

Recent numerical studies of the general relativistic (GR)
hydrodynamics in three spatial dimensions of close neutron star
binaries (NSBs) have uncovered evidence for the compression and
heating of the individual neutron stars (NSs) prior to
merger\cite{jay:wmm96,jay:mw97}.  This effect will have
significant effect on the production of gravitational waves, neutrinos
and, ultimately, energetic photons.  The study of the production of
these photons in close NSBs and, in particular, its
correspondence to observed GRBs is the subject of this
paper.

The gamma-rays arise as follows.  Compressional heating causes the
neutron stars to emit neutrino pairs which, in turn, annihilate to
produce a hot electron-positron pair plasma.  This pair-photon plasma
expands rapidly until it becomes optically thin, at which point the
photons are released.  We show that this process can indeed satisfy
three basic requirements of a model for cosmological gamma-ray
bursts:\\ 1) sufficient gamma-ray energy release ($> 10^{51}$ ergs) to
produce observed fluxes,\\ 2) a time-scale of the primary burst
duration consistent with that of a ``classical'' GRB ($\sim$ 10
seconds),\\ 3) peak of photon number spectrum matches that of
``classical'' GRB ($\sim 300$ keV).
\end{abstract}

\section*{Neutron Star Heating}

The method for solving the general relativistic field equations in
three spatial dimensions has been discussed in
\cite{jay:wmm96,jay:mw97}.  At each time slice we obtain an exact (to
numerical accuracy) instantaneous solution to the GR field equations.
The hydrodynamic equations are then evolved for the moving matter
against these GR fields.  This method ignores gravitational waves,
however in \cite{jay:wmm96} it was shown that the effect is very
small; ${\dot{J} \over \omega J} \sim 10^{-4}$, where $J$ is the
angular momentum and $\omega$ is the angular frequency of the NSB.

\begin{figure}
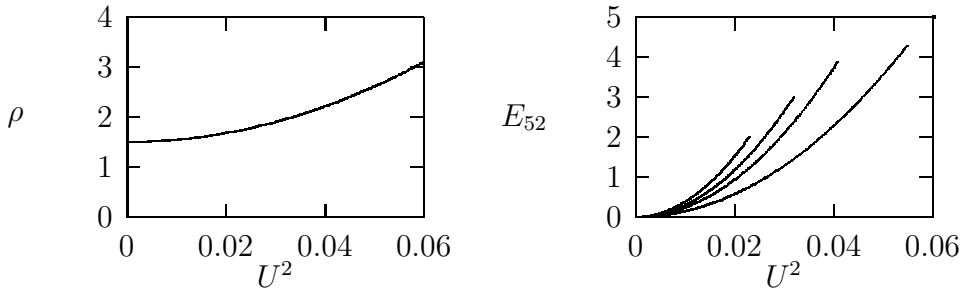

\begin{center}
\input{plot1}
\input{plot2}
\caption{Plots of central density $\rho\ (\times 10^{15} gm/cc)$ and
released gravitational energy $E_{52}\ (\times 10^{52} ergs)$ as a
function of $U^2$.  The $E_{52}$ plot shows a range of star masses:
(from left to right) 1.50, 1.45, 1.40, 1.35 $M_\odot$.  The EOS used
for these examples has a critical mass $M_{crit} = 1.64 M_\odot$.}
\label{jay:fig1}
\end{center}
\end{figure}

The computational evolution calculation of NSBs and their GR fields
begins by generating two identical non-spinning neutron stars with an
initial mass and an EOS.  The stars are allowed to evolve until a
stationary orbit is achieved with a prescribed initial angular
momentum.  A variety of systems have been studied over a range of star
masses, equations of state and initial angular momentums.  The key
result to report here is that the proper baryonic density of the stars
was observed to increase prior to the stars reaching their last stable
orbit.  This compression and heating can be parameterized in terms of
$U^2$, the squared amplitude of the spatial components of the
4-velocity (Figure \ref{jay:fig1}).  See
\cite{jay:mw97,jay:mw297,jay:mw397} for discussion of why the
compression is not adiabatic.

In \cite{jay:mw97} it is argued that the gravitational binding energy
will be converted into thermal energy.  This thermal energy will be
radiated via neutrino luminosity.  The time scale for the energy
emitted up to collapse $(t < 0)$ can be estimated from the gravitation
wave emission
\begin{equation}
E(t) = { E_{tot} \over [1 - (64/5) m^{5/3} \omega^{8/3}_0 t]^{1/2}}
\label{jay:E:energy}
\end{equation}
where $\omega_0$ is the final angular orbital velocity.

Some of the $\nu\overline{\nu}$ pairs emanating from the surface of a
hot neutron star will annihilate to create $e^+e^-$ pairs.  In order
to calculate an estimate of the efficiency of this process, the
numerical supernova model of Mayle \&
Wilson\cite{jay:wilsonmayle93,jay:mw97} was used.  It was found that
the $\nu + \overline{\nu} \rightarrow e^+ + e^-$ reaction is 3\%
efficient at the end of a standard supernova calculation when the
neutron star at the center is at its hottest and most compact.  This
simulation includes all GR effects, except the neutrinos are assumed
to travel in straight lines.  In the strong field environment around a
NSB the neutrino trajectories will be significantly bent, thus
increasing the chances of $\nu\overline{\nu}$ annihilation.  New
estimates show that this bending will augment $\nu\overline{\nu}$
annihilation by a factor of $\sim 3$.  So the efficiency of $\nu +
\overline{\nu} \rightarrow e^+ + e^-$ is estimated to be $\sim 10$\%.
It is also found that the entropy per baryon of the pair plasma is
greater than $10^{10}$ so relatively very few baryons are liberated
from the surface of the star.

From Figure \ref{jay:fig1} we see that a typical star emanates $\sim 3
\times 10^{52}$ ergs of gravitational energy as $\nu\overline{\nu}$
pairs.  Thus we have $\sim 6 \times 10^{52}$ ergs in
$\nu\overline{\nu}$ pairs from both stars.  A 10\% efficiency in $\nu
+ \overline{\nu} \rightarrow e^+ + e^-$ gives us $\sim 6 \times
10^{51}$ ergs of energy in the form of a $e^+e^-$ pair plasma-photon
gas.  The temperature of this gas near the NS is several MeV.

\section*{The Gamma Ray Burst}
Having roughly defined the initial parameters of the hot $e^+e^-$ pair
wind blowing off of a NS, we wish to follow its evolution and
characterize the observable gamma-ray emission.  It is important to
note that there are no free parameters in this model, barring
uncertainties in understanding and correctly calculating the physics;
our signature either corresponds to an experimentally observed
phenomenon (i.e. GRBs) or it does not.

The expanding $e^+e^-$ pair plasma emanating from a NS is modeled as a
spherically symmetric special relativistic fluid by the following
hydrodynamic equations:

\begin{equation}
{\partial D \over \partial t} = - {1 \over r^2} { \partial \over
\partial r} (r^2 D V)
\end{equation}
\begin{equation}
{\partial E \over \partial t} = - {1 \over r^2} { \partial \over
\partial r} (r^2 E V) - P \biggl[ {\partial \gamma \over \partial t} +
{1 \over r^2} {\partial \over \partial r} ( r^2 \gamma V) \biggr]
\end{equation}
\begin{equation}
{\partial S \over \partial t} = - {1 \over r^2} { \partial \over
\partial r} (r^2 S V) - {\partial P \over \partial r}
\end{equation}
where the radial 4-velocity U, Lorentz factor $\gamma$ and coordinate
velocity V are defined as
\begin{equation}
U \equiv {S \over D + \Gamma E},\ \ \  \gamma \equiv \sqrt{ 1 + U^2},\ \ \  V
\equiv {U \over \gamma}.  \label{jay:E:defu}
\end{equation}
D and E are the coordinate densities of baryons and thermal energy
($e^+e^-$ and photons) respectively.

The total energy equation, including photons and $e^+e^-$ pairs, is
\begin{equation}
E_{tot} = a T^4 + E_{pairs} (T).
\end{equation}

To track the $e^+e^-$ pairs we define a pair equation
\begin{equation}
{\partial N_{pairs} \over \partial t} = - {1 \over r^2} { \partial
\over \partial r} (r^2 N_{pairs} V) + \overline{\sigma v} N_{pairs}
(N_{pairs}^0 (T) - N_{pairs})/\gamma^2
\end{equation}
where the coordinate pair number density is $N_{pair}$ and
$\overline{\sigma v}$ gives the mean pair annihilation rate.
$N_{pairs}^0 (T)$ is given by a Fermi integral with a chemical
potential of zero.


To model the material blown off the surface of a NS we inject baryon
and pair-photon energy densities into the innermost zone at a rate
determined by the time derivative of the heating energy given in
Equation (\ref{jay:E:energy}).

The hydrodynamic equations are evolved, allowing the plasma to expand.
Once the system becomes transparent to Compton scattering, assuming
no further scattering, the calculation is stopped and the photon signal
is analyzed.  In the results presented here we have set $E_{tot} =
10^{51}$ ergs.  Since the entropy per baryon of the wind is quite high
we define the rate of injection of baryons as $\dot{D} = 10^{-10}
\dot{E}$.


Since the photons and $e^+e^-$ pairs appear to decouple at virtually
the same time throughout the entire fireball (radius $\sim 10^{12}$ cm),
we take this event to be instantaneous and to occur when the cloud
becomes optically thin to Compton scattering.  We then look at two
observables, the time integrated number spectrum $N(\epsilon)$ and
the total energy received as a function of time E(t).

\begin{figure}
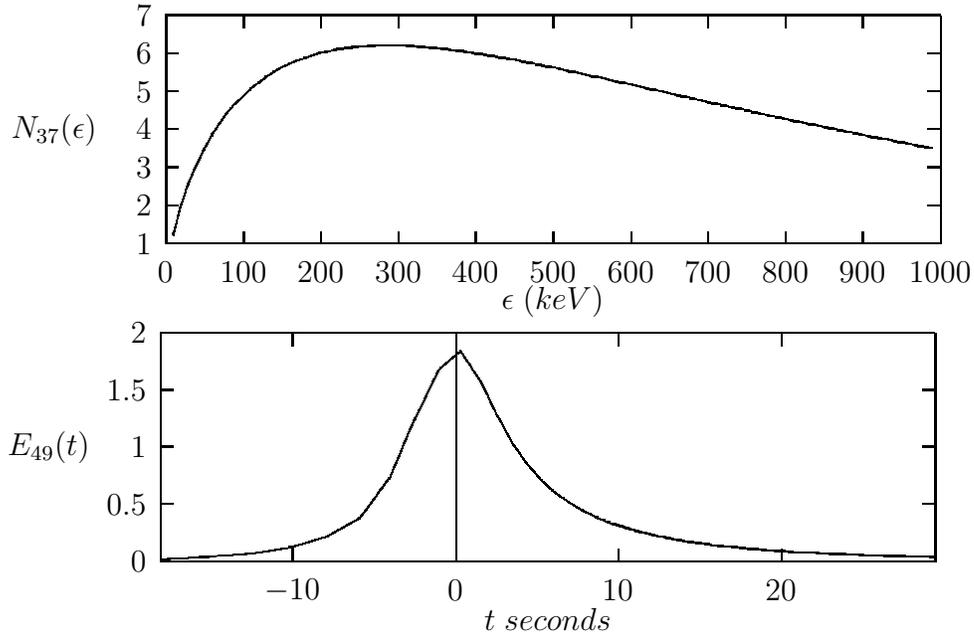

\begin{center}
\input{plot3}
\input{plot4}
\caption{({\bf Top}) Photon number spectrum $N_{37}(\epsilon)\ (\times
10^{37} photons/keV/4\pi )$ from $e^+e^-$ pair plasma
fireball. ({\bf Bottom}) Light curve $E_{49}(t)\ (\times 10^{49}
ergs/second/4\pi )$}
\label{jay:fig2}
\end{center}
\end{figure}

To get the spectrum, as mentioned above, we assume that the $e^+e^-$
pairs and photons are in thermodynamic equilibrium when they decouple.
Thus the photons in the fluid frame (denoted with a prime) make up a
Plank distribution.  We calculate the observed number spectrum, per
photon energy $\epsilon$, per steradian, of a relativistically
expanding spherical shell with radius R, thickness dR in cm, velocity
v where c=1, Lorentz factor $\gamma$ and fluid-frame temperature $T'$
to be
\begin{equation}
N(\epsilon) = 4\pi R^2 dR {\epsilon T' \over v \gamma} log
\Biggl[ {1 - exp[- \gamma \epsilon (1 + v)/T' ] \over 1 -
exp[ - \gamma \epsilon (1 - v)/T' ] } \Biggr]
\end{equation}
which has a maximum at $\epsilon_{max} \cong 1.39 \gamma T'\ eV$ for
$\gamma \gg 1$.  We may then sum this spectrum over all shells of our
fireball to get the total spectrum shown in Figure \ref{jay:fig2} ({\bf Top}).
Since we a priori assume the photons are thermal, our spectrum has a
high frequency exponential tail.  


A key feature of this spectrum is that its peak agrees with
observation.  It is interesting to note that the bulk of the photons
have a fluid frame temperature of only $\epsilon \sim 5-15$ eV, but
are Lorentz boosted by $\gamma \sim 10^4-10^5$.  Thus our spectrum
derives from a more relativistic fluid than other models.  The photons
to be observed at early times have about twice the energy of the later
photons.

To acquire the observed light curve $E(t)$ we consider two effects.
First is the relative arrival time of the first light from each shell.
Second is the shape of the light curve from a single
shell\cite{jay:fenimore}.  We find that, for our Plank distribution of
photons, a relativistically expanding shell of radius R will have a
time profile $E(\tau > R/c) \sim ({R \over c\tau})^4$.


In Figure \ref{jay:fig2} ({\bf Bottom}) we see an example of E(t) for NSB of
equal star mass.  Variation in the ratio of star mass in the NSB
effects the relative compression and heating rate of each star, thus
allowing a variety of GRB durations.  We estimate a range of burst
durations from several seconds to a few 10s of seconds.

Grant support is NSF PHY-97-22086, PHY-9401636 and DOE W-7405-ENG-48.


\begin{references}
\bibitem{jay:wmm96}Wilson, J.\ R., Mathews, G.\ J. \& Marronetti, P.,
{\it Phys.\ Rev.\ D.}\ {\bf 54}, 1317 (1996).
\bibitem{jay:mw97}Mathews, G.\ J., \& Wilson, J.\ R., {\it
Astrophysical J.} {\bf 482}, 929---941, (1997).
\bibitem{jay:mw297}Wilson, J.\ R., \& Mathews, G.\ J., submitted {\it
Phys.\ Rev.\ D} (1997).
\bibitem{jay:mw397}Mathews, G.\ J., submitted {\it Phys.\ Rev.\ Lett.} (1997).
\bibitem{jay:wilsonmayle93}Wilson, J.\ R., \& Mayle, R.\ W., {\it
Phys.\ Rep.} {\bf 227}, 97, (1993).
\bibitem{jay:fenimore}Fenimore, E.\ E.\ et al., {\it Astrophysical
J.\ } {\bf 473}, 998, (1996).
\end{references}
\end{document}